\documentstyle [11pt]{article}
\def\journal#1#2#3#4{\ {\sl #1}\ \underline{\bf #2}, {#3}\  ({#4})}

\def\ActaPolonica{\journal{Acta \ Physica \ Polonica\/}}

\def\AnnPhys{\journal{Ann.\ Phys.}}

\def\ibid{\journal{\em ibid.}}
\def\IJMPA{\journal{Int.\ J.\ Mod.\ Phys \ {\bf A}}}

\def\NPB{\journal{Nucl.\ Phys.\ {\bf B}}}

\def\PLB{\journal{Phys.\ Lett.\ {\bf B}}}
\def\PhysRev{\journal{Phys.\ Rev.}}

\def\PRD{\journal{Phys.\ Rev.\ {\bf D}}}
\def\PRL{\journal{Phys.\ Rev.\ Lett.}}

\def\SovJNuclPhys{\journal{Sov.\ J.\ Nucl.\ Phys.}}

\def\ZPhys{\journal{Z.\ Phys.}}

\begin{document}

\newcommand{\be}{\begin{equation}}
\newcommand{\ee}{\end{equation}}
\newcommand{\barray}{\begin{eqnarray}}
\newcommand{\earray}{\end{eqnarray}}
\newcommand{\gn}{\mbox{$\gamma_{\stackrel{}{5}}$}}
\newcommand{\adag}{a^{\dagger}_{p,s}}
\newcommand{\atildedag}{\tilde{a}^{\dagger}_{-p,s}}
\newcommand{\bdag}{b^{\dagger}_{-p,s}}
\newcommand{\btildedag}{\tilde{b}^{\dagger}_{-p,s}}
\newcommand{\adagbdag}{a^{\dagger}_{p,s} b^{\dagger}_{-p,s}}
\newcommand{\aps}{a^{}_{p,s}}
\newcommand{\bps}{b^{}_{-p,s}}
\newcommand{\Adag}{A^{\dagger}_{p,s}}
\newcommand{\Bdag}{B^{\dagger}_{-p,s}}
\newcommand{\Aps}{A^{}_{p,s}}
\newcommand{\Bps}{B^{}_{-p,s}}
\newcommand{\ApL}{A^{}_{p,L}}
\newcommand{\BpL}{B^{}_{-p,L}}
\newcommand{\ApR}{A^{}_{p,R}}
\newcommand{\BpR}{B^{}_{-p,R}}
\newcommand{\BdagL}{B^{\dagger}_{-p,L}}
\newcommand{\BdagR}{B^{\dagger}_{-p,R}}
\newcommand{\eps}{\epsilon}
\newcommand{\go}{\left( \gamma_o - \vec{\gamma} \cdot \hat{n} \right)}
\newcommand{\abab}{a^{\dagger}_{p,L}\,b^{\dagger}_{-p,L}\,a^{\dagger}_{p,R}
                   \,b^{\dagger}_{-p,R}}
\newcommand{\alphai}{\alpha_{i}}
\newcommand{\limit}{\lim_{\Lambda^2 \rightarrow \infty}}
\newcommand{\prodps}{\prod_{p,s}}
\newcommand{\prodp}{\prod_{p}}
\newcommand{\sumps}{\sum_{p,s}}
\newcommand{\psibar}{\bar{\psi}}
\newcommand{\psibarpsi}{ < \bar{\psi} \, \psi
            > }
\newcommand{\PsibarPsi}{ < \bar{\Psi} \, \Psi
            > }
\newcommand{\psibeta}{\psi^{}_{\beta}}
\newcommand{\psibarbeta}{\bar{\psi}_{\beta}}
\newcommand{\psidag}{\psi^{\dagger}}
\newcommand{\psidagbeta}{\psi^{\dagger}_{\beta}}
\newcommand{\psiL}{\psi_{_{L}}}
\newcommand{\psiR}{\psi_{_{R}}}
\newcommand{\Q}{Q_{_{5}}}
\newcommand{\Qa}{Q_{_{5}}^{a}}
\newcommand{\Qbeta}{Q_{5}^{\beta}}
\newcommand{\qqbar}{q\bar{q}}
\newcommand{\TH}{\Theta}
\newcommand{\thetap}{\theta_{p}}
\newcommand{\costhetap}{\cos{\thetap}}
\newcommand{\sinthetap}{\sin{\thetap}}
\newcommand{\thetaset}{\{ \thetap  \}}
\newcommand{\thetapi}{\thetap{}_{i}}
\newcommand{\Tomega}{\frac{\Tprime}{\omega}}
\newcommand{\pomega}{\frac{p}{\omega}}
\newcommand{\Tprime}{T^{'}}
\newcommand{\vac}{| vac \rangle}
\newcommand{\vacbeta}{| vac \rangle_{_{\beta}}}
\newcommand{\x}{\vec{x},t}
\newcommand{\xPrime}{\vec{x} - \hat{n} ( t - t'), t'}
\newcommand{\y}{\vec{y}, y_o}

\thispagestyle{empty}

\vspace*{-.15in}

\hspace*{\fill}\fbox{CCNY-HEP-93-15}

\begin{center}
{\Large {\bf Braaten-Pisarski Action, Disoriented Chiral Condensate, and
             Chiral Symmetry Non-Restoration  }
\fnsymbol{footnote}\footnote
{\samepage \sl
\noindent \parbox[t]{138mm}{ \noindent
                       Contributed talk at the 3rd Thermal Fields Workshop
                       held Aug 16 - 27, 1993 at Banff, Canada. \\
                       Parts of this work have been supported in part by a
                       grant from NSF and from PSC-BHE of CUNY.
}
}
}\\
\baselineskip 5mm
\ \\
Ngee-Pong Chang (npccc@cunyvm.cuny.edu)\\
Department of Physics\\
City College \& The Graduate School of City University of New York\\
New York, N.Y. 10031\\
\  \\
Aug 23, 1993  \\
\end{center}

\vspace*{-.25in}

\noindent\hspace*{\fill}\parbox[t]{5.5in}{
        \hspace*{\fill}{\bf Abstract}\hspace*{\fill} \\

        \em
        The QCD effective action at high $T$ shows a manifest
        global chiral symmetry.  And calculations show that
        the order parameter $\psibarpsi$ vanishes above $T_c$. \\
        \hspace*{\fill}{\em Is this not evidence for chiral symmetry
        restoration?}\hspace*{\fill}\\

        It has been popular to refer to this $T_c$ as chiral symmetry
        restoration temperature because it fits into our
        prejudice that chiral symmetry is like an `ordered' state, and
        at high $T$ it must become disordered.  In fact,
        NJL$\cite{NJL}$ ground state is not an ordered spin state.
        I give an example of a generalized chiral broken NJL ground state
        for  which $\psibarpsi$ nevertheless vanishes. \\

        The recent scenario of a generic class of disoriented
        chiral condensate$\cite{dcc}$ offers another example where
        $\psibarpsi$ in each little domain is nonzero, but the average
        over all space of $\psibarpsi$ vanishes.  Such a dcc ground state
        continues to break chiral invariance. \\

        But how do you reconcile this with the apparent chiral
        symmetry at high $T$? \\

        The Braaten-Pisarski action$\cite{BP}$ is a good laboratory to
        investigate the subtleties of high temperature chiral symmetry.
        By carrying out a canonical quantization of
        this highly nonlocal action, I demonstrate how the
        thermal vacuum at high $T$ conserves the new $\beta$-chirality
        but breaks the old $T=0$ chirality. \\

        Lattice calculations show that the pion develops a
        screening$\cite{screening mass}$ mass at high $T$.  Our continuum
        field theory calculations$\cite{Chang-QCD-pion}$
        show that the QCD pion remains massless for all $T$.
        I conclude the talk by showing how the hot pion manages to
        accomodate the two results by propagating in the early
        universe with a halo.
                              }\hspace*{\fill} \\


\section{Introduction}

        In these idyllic surroundings and cool environment of Banff, I am
        pleased to be discussing with you today a different view of the
        phase transition that takes place at a much hotter temperature,$T_c$.
        I am referring to the vanishing of $\psibarpsi$ for $T$ above $T_c$.
        As mentioned in the abstract, I will give in this talk the
        background to my contention why in spite of appearances, {\em the
        chiral symmetry that we know at $T=0$ is \underline{not }
        restored at high $T$}.

        And yet the QCD effective action at high $T$ is manifestly
        chiral invariant.  Is chiral symmetry not restored?
        The Braaten-Pisarski effective action$\cite{BP}$ is a good
        laboratory in which to point out the subtleties involved.
        It is globally chiral invariant, so that it would appear to be
        consistent with the popular notion of chiral restoration.
        And yet the quark propagates through a hot
        medium$\cite{Weldon-Klimov}$
        as if it has a pseudo-Lorentz invariant
        mass$\cite{Donoghue-Chang-hiT-Barton}$, $\Tprime$ ($\equiv
        g_r\sqrt{ C_f} T/2 $ ).
        While chiral symmetry at $T=0$ requires that the fermion
        be massless, this new chiral symmetry at high $T$
        allows for what has been referred to as thermal mass.
        This thermal mass is there for both QED and QCD.

        The Noether charge associated with the high temperature chiral
        phase is demonstrably different than the usual Noether charge.
        I have performed a canonical quantization of the BP effective
        action for the quark field, and will present to you the
        canonical expansion for the two Noether charges, so that
        you can judge for yourself.

        The implications of a continued breaking of chiral symmetry
        at high $T$ are, of course, quite astounding.  If you believe
        in a fundamental Higgs field for the Standard Model, then
        above the (very high) electroweak transition temperature,
        the vev for the Higgs field vanishes, and the quarks no longer
        acquire a mass through the Yukawa coupling.
        The Nambu-Goldstone theorem now forces
        the pion {\em to be strictly massless in the early universe} even
        in the presence of electroweak interactions.

        In the concluding section of my talk, I present a picture
        of the propagation of the pion through the early hot medium,
        and show how it propagates with light velocity, but acquires
        a halo.

        Presumably, the presence of the $\qqbar$ bound state in the
        early universe will have some new and subtle effect. But
        as to what that will be, I can only hope someone in the audience
        will be expert enough to advise me.

\section{NJL ground state at $T=0$}

        At zero temperature, the pioneering
        work of Nambu and Jona-Lasinio$\cite{NJL}$ has taught us how
        massless fermions manage nevertheless to acquire dynamical mass.
        The NJL ground state is made up of massless quark-antiquark pairs
        with the same helicity
\be
        \vac  =  \prodps \left( \cos{\thetap} \; - s \;
                 \sin{\thetap} \adagbdag \right)
                 | 0 \rangle                           \label{NJL}
\ee
        where $s$ is defined to be $\pm 1$ respectively for $R$ and $L$
        helicities.
        The observed massive quarks are the quasi-particle excitations off
        this ground state.
\barray
        \Aps \,\;  &=&     \cos{\thetap} \;\aps \,\; + s \; \sin{\thetap}
                                         \;\bdag   \label{Aps} \\
        \Bps  &=&     \cos{\thetap} \,\bps - s \; \sin{\thetap}
                                         \;\adag   \label{Bps}
\earray
        The mass gap associated with these quasi-particles
        are directly related to the amount of $q \bar{q}$  mixing in the
        NJL vacuum
        ($\tan{2\thetap}   =   m/p $ )
        and may be determined self-consistently from the
        dynamics by solving the famous gap equation.

        The original scale invariant Lagrangian is formally invariant
        under the global chiral transformation
\be
        \psi (\x)  \rightarrow   {\rm e}^{i \alpha \gn }
                                 \; \psi (\x)         \label{chi-transf}
\ee
        The Noether charge $\Q$ that generates the phase changes
        for the massless quark and antiquark operators is given by
\be
        \Q     =   \frac{1}{2} \;\;\int d^3 x \; \psi^{\dagger} \, \gn \,
                               \psi
               =   - \frac{1}{2} \; \sumps \; s \; \left(
                     \adag \aps + \bdag \bps
                     \right)                          \label{Q5}
\ee
        and you can see how the NJL ground state, eq.(\ref{NJL}), is
        not annihilated by this $\Q$.

        A signature for this spontaneous breakdown is the nonvanishing
        of the order parameter, $\psibarpsi$.  QCD sum rules, as well
        as lattice calculations and continuum field theory have all
        demonstrated this.  Our earlier calculation$\cite{Chang-PRL}$
        showed the connection between dynamical symmetry breaking
        and bifurcation theory, and led to a universal prediction
\be
        \psibarpsi   =  - 0.0398 \,N_c \,N_f \;\Lambda_c^3
\ee
        Associated with this breakdwon is the
        presence of the Nambu-Goldstone pion in $T=0$ QCD.  If we
        could ignore electroweak interactions, then the pion is
        to be massless.  Because of the fermion Yukawa couplings,
        the electroweak breaking feeds a tree level mass to the
        quarks, and the QCD pion is no longer massless, and
        acquires the observed $135$ $MeV$.

        As $T$ increases, it has been observed that $\psibarpsi$
        vanishes at some $T_c$, and stays zero for $T$ above it.
        Our calculations$\cite{Donoghue-Chang-hiT-Barton}$ show this $T_c$
        to be $\Lambda_c {\rm e}^{2/3}$.
        In the popular folklore this phase transition is interpreted
        as chiral symmetry
        restoration at high temperatures, in line with what happens, say,
        with the Heisenberg ferromagnet.

        It does not have to be so.
        I quote here a very simple `counterexample' of a generalized
        NJL ground state that would have $\psibarpsi = 0$, and yet
        is manifestly not chiral invariant.  Namely, put a phase
        factor $i$ in the NJL ground state
\be
        \vac ' =  \prodps \left( \cos{\thetap} \; - \; i\; s \;
                 \sin{\thetap} \adagbdag \right)
                 | 0 \rangle                           \label{NJL'}
\ee
        It is an instructive but very simple exercise to check that
        $\psibarpsi$ vanishes with respect to this new ground state.
        I mean this example to show that $\psibarpsi$ is an incomplete
        order parameter for chiral symmetry breaking.  It measures
        the `real' part of the NJL ground state, and misses out on the
        `imaginary' part of a generalized NJL state.\footnote{
\samepage \sl
        The full extent of the chirality algebra and the larger
        set of order parameters associated with the algebra will be
        described in a forthcoming separate paper.
                                                              }

\section{Disoriented Chiral Condensate}
        To be sure, the vanishing $\psibarpsi$ at $T_c$ signals a phase
        transition.
        If it is not chiral symmetry restoration, then what could it
        be?    I would like to suggest to you that it is a transition to
        the new $\beta$-chiral phase, and the generic class of disoriented
        chiral condensate$\cite{dcc}$  ({\em dcc})\footnote{
\samepage \sl      Since R. Pisarski has already covered it in his lectures
                   here at the workshop, I will conserve space and
                   not describe the model here in any detail.
                                              }
        is a good realization of this new equilibrium phase.

        In this scenario, the ground state (i.e. the universe)
        above $T_c$ breaks up into many little domains, inside each of which
        $\psibarpsi$ takes a different value, so that when
        averaged over all domains, $\psibarpsi$ vanishes.  Under a global
        chiral transformation, each domain would further tilt, so that
        in such a scenario, the class of disoriented chiral condensate
        vacua is not invariant under the $\alpha$-chiral
        transformation of eq.(\ref{chi-transf}).

\section{Braaten-Pisarski Action}

        Because I am focussing on the chiral symmetry aspects, I will
        proceed forthwith to consider only the two fermion sector
        and set the background gluon field to zero. The form of
        the BP action\footnote {
\samepage \sl
        In real time formalism, there is a contribution to the full
        action from the $\tilde{\psi}$ associated with the heat bath.
        It is given by the {\em tilde} operation acting on the
        action in eq.(\ref{BP}), such that $I_{\rm full}
        = I_{BP} \{ \psi \} - I_{BP} \{ \tilde{\psi} \}$
                               }
        that we shall study then has the form
\be
        I_{\rm BP} = \int d^4 x \left\{
                  - \psibarbeta \gamma_{\mu} \partial^{\mu} \psibeta
                  - \frac{\Tprime {}^2}{4} \int dt' \left< \psibarbeta (\x)
                  \left( \gamma_o - \vec{\gamma} \cdot \hat{n} \right)
                  \psibeta(\xPrime) \right> \eps (t-t')
                             \right\}                 \label{BP}
\ee
        where the angular bracket denotes an average over the orientation
        $\hat{n}$.

        The second term in the BP action leads to the thermal mass for
        the fermion.  The BP appears to be chiral invariant under the
        transformation
\be
        \psibeta (\x ) \rightarrow {\rm e}^{i \beta \gn } \; \psibeta (\x )
\ee
        The Noether charge$\cite{Weldon-BP}$ associated with this chirality
        is however not given by eq.(\ref{Q5}) but\footnote{
\samepage \sl    Likewise, the complete charge includes an identical but {\em
        negative} contribution where $\psi$ everywhere has been replaced
        by $\tilde{\psi}$ field.
                                               }
\barray
       \Qbeta &=&  \frac{1}{2} \;\; \int d^3 x \left\{ \; \psidagbeta
                        \, \gn \,
                        \psibeta - \frac{\Tprime {}^2}{8} \int dt_1 dt_2
                                   \; \epsilon (t_1 -t) \epsilon (t -t_2)
                                   \right.  \nonumber \\
             &&     \left. \, \psidagbeta (\vec{r} + \frac{\hat{n}}{2}
                                    (t_1 - t_2), t_1) \,
                    \left(  1 + \gamma_o \vec{\gamma} \cdot \hat{n} \right)\gn
\
                                     \psibeta (\vec{r} - \frac{\hat{n}}{2}
                                     (t_1 - t_2), t_2)
                                     \displaystyle
                                     \right\}    \label{Qbeta}
\earray
        To understand further the physics of this Noether charge, it is
        necessary to quantize the BP action.

        The BP action is manifestly nonlocal.  There is an essential difference
        between this nonlocality and the situation at $T=0$.
        There the nonlocality is weak, since they are protected by appropriate
        powers of the cutoff $\Lambda$ in the denominator.  A derivative
        expansion thus makes sense if one is talking about physics at
        a momentum scale below the cutoff.  For high $T$, however, the
        nonlocality is proportional to $T^2$ {\em in the numerator}, and no
        derivative expansion is possible.

        Because of the nonlocality, $\psibeta$ is not a canonical field
        in the BP action.\footnote{  \label{rel-Heisen}
\samepage \sl   Note that $< T( \psibeta (\x) \psibarbeta (0) )>_{\beta}$
                is by definition also equal to the thermal average of
                the Heisenberg fields $ \;\; \sum_{n} \;\;<n| T( \psi(\x)
                \psibar (0) ) |n> {\rm e}^{- \beta E_n} / Z $.
                Therefore it is reassuring to verify that the vacuum
                expectation  $< \{ \psibeta , \psidagbeta \} > =
                \delta (\vec{x} - \vec{y}) $, even though $\psibeta$
                does not satisfy it as an operator identity.
                                  }
        To quantize this action, it is convenient
        to work in momentum space, where the action takes the form
\be
        I_{\rm BP} = \int d^4 p \left\{
                      - i \psibarbeta (p) \left( \vec{\gamma} \cdot \vec{p}
                      - \gamma_o p_o \right) \psibeta (p)
                      + i \frac{\Tprime \stackrel{{}^2}{}}{2 \;\;}
                      \psibarbeta (p) \left( \vec{\gamma}\cdot \vec{p} \; a
                      - \gamma_o p_o \; b \right) \psibeta (p)
                                 \right\}
\ee
        with $a \equiv \frac{p_o}{2p^3} \left| \frac{p_o + p}{p_o - p}\right|
        - \frac{1}{p^2} $ and $b \equiv \frac{1}{2p p_o} \left| \frac{p_o + p}
        {p_o - p}\right| $

        It may be checked that this action gives rise to the fermion
        propagator $< T( \psibeta (\x) \psibarbeta (0)) >_{\beta}$
        (see footnote \ref{rel-Heisen}) with the usual analyticity properties,
        {\em viz.} positive and negative energy poles from {\em both}
        particles and holes of mass $\Tprime$,
        plus a parallel pair of conjugate plasmino cuts in $p_o$ plane that
        extend from $-p$ to $p$ just above and below the real $p_o$ axis.
        The discontinuity across each cut is of order $\Tprime {}^2$.

        For our discussion here, we shall work to order $\Tprime$ and
        ignore the contributions due to the plasmino cut.
        The canonical field $\Psi$ may be obtained by a redefinition
        of $\psi$
\be
        \psi (p)  =  {\rm e}^{i \frac{\Tprime}{2} \TH} \; \Psi (p)
                     \sqrt{z_p}
\ee
        where $\TH \equiv \vec{\gamma}\cdot \vec{p} \;a \; - \gamma_{o} p_o
        \; b $, and $z_p$ is the wave function renormalization, which
        to order $\Tprime$ is simply unity.  With this
        field redefinition, we find
\be
        I_{\rm BP} = \int d^4 p \left\{
                      - \; i \; \overline{\Psi} (p) \left( \vec{\gamma}
                                \cdot \vec{p}
                      - \gamma_o p_o \right) \, \Psi (p)
                      - \Tprime \; \overline{\Psi}(p) \Psi (p)
                                 \right\}
\ee
        confirming that $\Psi$ indeed is the canonical massive Dirac field
        for the BP action. To specify the $\pm i\eps$ boundary conditions,
        we require that at $t=0$, $\Psi$ field coincides with the free
        massless $\psibeta$ field in a thermal equilibrium,
        so that equations (\ref{Aps},\ref{Bps}) hold with mass
        gap equal to $\Tprime$, and likewise for the tilde degrees of
        freedom. If we work with the Bogoliubov transformed basis
\be
        | 0 \rangle_{_{\beta}} = \prodps
        \left( \frac{1}{\sqrt{2}}-  \frac{1}{\sqrt{2}} \adag \atildedag \right)
        \left( \frac{1}{\sqrt{2}}+  \frac{1}{\sqrt{2}} \bdag \btildedag \right)
                                   | 0 \rangle
\ee
        the new vacuum is a generalized NJL vacuum
\be
        | vac \rangle_{_{\beta}} = \prodps
        \left( \cos{\thetap} - \;s \, \sin{\thetap} \;
                   a_{\beta,p,s}^{\dagger} b_{\beta,-p,s}^{\dagger} \right)
        \left( \cos{\thetap} - \;s \, \sin{\thetap} \;
               \tilde{a}^{\dagger}_{\beta,p,s}
               \tilde{b}^{\dagger}_{\beta,-p,s} \right)
                                   | 0 \rangle_{_{\beta}}
\ee
        where $a_{\beta,p,s}^{\dagger},b_{\beta,p,s}^{\dagger} $ are
        the Bogoliubov transform of the usual massless operators.

        The full Noether charge $\Qbeta$ may now be expressed in terms of the
        canonical annihilation and creation operators of the massive
        Dirac field as
\be
        \Qbeta {}_{\rm full}   =  - \frac{1}{2} \;\sumps \; s \; \left(
                      \Adag \Aps + \Bdag \Bps
                      - \tilde{A}^{\dagger}_{p,s} \tilde{A}^{}_{p,s}
                      - \tilde{B}^{\dagger}_{p,s} \tilde{B}^{}_{p,s}
                                  \right)
\ee
        If you compare this with the canonical expansion for the
        $T=0$ Noether charge, eq. (\ref{Q5}), you'll see why
        $\Q$ does not annihilate the generalized NJL vacuum, while the
        $\Qbeta$ (expressed in terms of the massive quasiparticle
        operators) does.

\section{Pion Halo}
        I conclude my talk by showing you how a QCD pion at high
        temperature can propagate as a massless particle and
        yet have a screening mass proportional to $T$.
        In the language of real time thermal field theory, it is easy
        to find an example of such a particle.  For the
        physical massless pole is determined from the condition that
        the denominator of the propagator vanish
\be
        \Gamma^{(2)}_{\pi} (p, p_o, T) = p^2 ( 1 + {\cal A} )^2 - p_o^2
                                     ( 1 + {\cal B} )^2 = 0
\ee
        where ${\cal A}$ and ${\cal B}$ are functions of $p, p_o, T$.
        The screening mass on the other hand comes from integrating over
        the $x, y, t$ coordinates (i.e. set $p_x = p_y = p_o = 0$)
        in the propagator, so that the pole for the correlation
        function in $z$ occurs at $p_z = i m_{sc}$, where
        $1 + {\cal A} ( im_{sc}, 0, T)  = 0$
        In terms of a physical picture, when we receive light from a
        charged particle, we see it at its retarded position, and
        it is a sharp image.  For the pion, the retarded function
        reads
\be
        D_{\rm ret} (\x) = \theta (-t) \left\{ \delta(t^2 - r^2)
                    + \frac{T}{r} \theta(t^2 - r^2)
                    \left[ {\rm e}^{-T | t-r| }
                    +  {\rm e}^{-T | t+r| } \right] \right\}
\ee
        so that the screening mass leads to an accompanying modulator
        signal that `hugs' the light cone, with a screening length
        $\propto 1/T$.



\vspace*{-.15in}

\begin{thebibliography}{99}
\bibitem{NJL}
              Y. Nambu and G. Jona-Lasinio, \PhysRev {122}
              {345} {1961}; \ibid {124} {246} {1961}.
\bibitem{dcc}
A. Anselm, M. Ryskin \PLB {226} {1991};
J.-P. Blaizot, A. Krzywicki \PRD {46} {246} {1992};
J. Bjorken, \IJMPA {7} {4189} {1987},
\ActaPolonica {B23} {561} {1992}; K. Kowalski, C.
Taylor preprint hepph/9211282 (1992); J. Bjorken, K. Kowalski, C.
Taylor SLAC preprint SLAC-PUB-6109 (1993).

\bibitem{BP}
              J.C.Taylor and S.M.H.Wong, \NPB{346}{115}{1990};
              E. Braaten and R. Pisarski, \PhysRev {D45}{1827}{1992};
              J. Frenkel and J.C. Taylor, \NPB{374}{156}{1992}.

\bibitem{screening mass}
              A. Gocksch and A. Soni, \ZPhys {53}{517} {1992}.

\bibitem{Chang-QCD-pion}
              L.N. Chang, N.P. Chang, \PhysRev {D45}{2988} {1992}.

\bibitem{Weldon-Klimov}
              H.A. Weldon, \PhysRev {D26} {2789} {1982}; V.V. Klimov,
              \SovJNuclPhys {33}{934} {1981}.
\bibitem{Donoghue-Chang-hiT-Barton}
              J.F. Donoghue and B.R. Holstein, \PhysRev  {D28}
              {340} {1983}; \ibid {29}{3004 (E)} {1984};
              L.N. Chang, N.P. Chang, K.C. Chou,
              \PhysRev {D43}{596} {1991}.  See also
              G. Barton, \AnnPhys {200}{271} {1990}.
              In contrast with ref. \cite{Weldon-Klimov}, the authors
              here take the perturbative approach and regard $\Tprime$ as
              a small parameter.
              The difference is negligible for $T >> p >> \Tprime$
              in the range where $p$ is still `soft'.

\bibitem{Chang-PRL} L.N. Chang and N.P. Chang,
              \PRL {54}{2407} {1985}.
      See also H. Munczek, \ZPhys {C32}{585} {1986}.
\bibitem{Weldon-BP} A.H. Weldon, Proceedings of Winnipeg Summer School 1992,
      Canadian J. Phys.  See also J.P. Blaizot, E. Iancu, {\em Soft
      Collective Excitations in Hot Gauge Theories}, SACLAY-SPHT-93-064,
      Jun 93.

\end{thebibliography}
\end{document}